# Precinct Size Matters - The Large Precinct Bias in US Presidential Elections

**G.F. Webb**
(Vanderbilt University, Nashville, TN USA)

**Abstract**

Examination of precinct level data in US presidential elections reveals a correlation of large precincts and increased fraction of Republican votes. The large precinct bias is analyzed with respect to voter heterogeneity and voter inconvenience as precinct size increases. The analysis shows that voter inconvenience is a significant factor in election outcomes in certain states, and may significantly disadvantage Democratic candidates.

**1. Introduction**

Precincts are the fundamental administrative voting units of US elections. A precinct has a designated polling location for precinct residents to vote, although sometimes a precinct has more than one polling place, or more than one precinct will vote in the same polling place. The number of voters in precincts varies considerably throughout the US. According to the federal US Election Assistance Commission, there were 174,252 precincts and 113,754 polling places in the 2004 US presidential election, with an average of approximately 800 voters per precinct. The availability of precinct level data also varies considerably throughout the US. Many states do not provide accessible precinct data at the state level, nor do counties at the county level. In the 2008 election, according to the Atlas Forum, 34 states provided precinct data, although not always in easily assessable format. In the study here precinct level data for 20 states is obtained either directly from state governments or from the Harvard Election Data Archive. The objective is to examine the role of precinct size on precinct vote outcome. Two possible correlations of precinct size and outcome will be analyzed. The first is the significance of voter heterogeneity as precinct size increases. The second is the significance of voter inconvenience as precinct size increases. The first is symmetric with respect to both political parties, but the second is not.

    The demographic and sociologic heterogeneity of voters varies greatly in precincts, and intuitively, heterogeneity increases as precinct size increases. A strong vote outcome for one side in a precinct might be diminished if the precinct were expanded to include a neighboring region with a very different demographic composition. This correlation of precinct size and vote outcome as a function of voter heterogeneity is symmetric with respect to both political parties. The impact of voter inconvenience also varies greatly in precincts, and again, intuitively, increases as precinct size increases. Here voter inconvenience is understood generally, and includes long lines, long wait times, uncomfortable waiting conditions, malfunctioning voting machines, understaffed poll workers, insufficient voting materials, excessive ballot length, complicated voting instructions, and other impediments to the voting process. This correlation of precinct size and vote outcomes as a function of voter inconvenience, however, is asymmetric with

respect to political parties. The economic and demographic factors that distinguish Democratic and Republican voters also distinguish their capacity to cope with voter inconvenience. Democratic voters often have more constrained voting schedules and more limited transportation options, and thus less flexibility in managing voting times. There is, in general, a greater loss of votes due to voter inconvenience for the Democratic candidate, than the Republican candidate.

## 2. Results

It is well known that the most populous counties throughout the US are biased toward Democratic candidates in presidential elections. It is less well known that large precincts throughout the US also have a presidential candidate bias, designated here as the large precinct bias (LPB). LPB means an advantage to the Republican (red) or Democrat (blue) presidential candidate as precinct size increases above a designated total precinct vote count. LPB does not mean that the red or blue candidate necessarily wins in large precincts, but rather that the relative vote fraction of the red or blue candidate increases as precinct size increases in these large precincts.

LPB varies considerably from region to region, state to state, and county to county. Nevertheless, LPB patterns can be detected in states with comparable demographic characteristics. To some extent it is surprising that such patterns do exist, since the distinction of small and large precincts is not readily correlated to political or demographic quantification. In most states precinct size varies considerably within demographic sectors, although urban areas do typically possess more large precincts than other areas, as evidenced in the Stanford Election Atlas. What is very surprising is that LPB usually tends to benefit the Republican presidential candidate, despite the correlations between precinct size, urban locations, and Democratic presidential candidate advantage in urban locations.

Precinct level data reveals LPB immediately by comparing the gain or loss of blue and red votes in precincts designated as large. This comparison, however, does not reveal relative LPB in precincts won by the blue candidate or precincts won by the red candidate. Either could increase or decrease relative to the other to explain the differential LPB of total votes cast in large precincts. The red fraction in large precincts could increase with precinct size, but is the increase due to a gain in the red fraction or a loss in the blue fraction or both? The separation of blue win precincts and red win precincts aids the understanding of LPB. If red fractions increase in large blue win precincts and blue fractions increase in large red win precincts in a comparable way, then the symmetry implies that LPB is due to shared red and blue demographic characteristics, such as increased voter heterogeneity in large precincts. If, however, red fractions increase in large blue win precincts, but blue fractions do not increase in large red win precincts, then the asymmetry implies that LPB is due to unequal demographic characteristics, such as increased voter inconvenience of blue voters.

The analysis here will reveal that in many states red LPB is significantly present in large blue win precincts and blue LPB is significantly absent in large red win precincts. This asymmetry is especially evident in large populous states. Because of the difficulties in accessing precinct level election data, the analysis is based on a limited number of states in the presidential elections of 2008 and 2012. Precinct level data was obtained directly from the state governments of Michigan, Ohio, Tennessee, and Wisconsin. Data for Alabama, Arizona, Colorado, Delaware, Florida, Idaho, Indiana, Kansas, Louisiana, Mississippi, Missouri, Nevada, New Mexico, North Carolina, Pennsylvania, and Washington was obtained from the Harvard Election Data Archive (data files of Stephen Ansolabehere and Jonathan Rodden - http://projects.iq.harvard.edu/eda/data).

## 3. Methods

The analysis of LPB is as follows:

(1) In each state precincts are separated into those won by the Democratic candidate (blue win) and those won by the Republican candidate (red win). The blue win precincts and red win precincts are each ordered with respect to precinct total votes from smallest to largest (allowing repeats). This ordering counts only actual vote totals for the Democrat and Republican presidential candidates (not other candidates nor write-ins), and neither the census population of the precinct nor the number of registered voters in the precinct. Distinction of absentee, early, and election-day polling are not made, since, in most cases, such data is not available at the precinct level.

(2) Let the blue win ordered precincts be labeled $\{(b_i + r_i)\}_{i=1}^{N}$, where $b_i$ ($r_i$) is the number of blue (red) votes in the $i^{th}$ precinct and $b_1+r_1$ ($b_N+r_N$) is the smallest (largest) precinct. Precincts with $b_i+r_i \geq 800$ are viewed as large precincts (the designation is arbitrary and could be adjusted higher or lower). Let $N^*$ be the smallest index such that $b_{N^*}+r_{N^*}$ is $\geq 800$. A least squares linear regression is fitted to the sequence of Republican vote fractions $\{r_i/(b_i + r_i)\}_{i=N^*}^{N}$ with slope denoted by *redSLOPE*. If *redSLOPE* > 0 (< 0), then the blue win ordered precincts have, on average, red vote fractions increasing (decreasing) in precincts with vote totals larger than 800. A similar framework is carried out for red win precincts.

(3) The value of *redSLOPE* can be used to measure the LPB for blue win ordered precincts. If *redSLOPE* > 0, then the gain in the red vote fraction in blue win precincts $\geq 800$ at the $i^{th}$ precinct in the ordering is $redSLOPE*(b_i+r_i – 800)$ (in terms of the linear regression, not the actual vote). The gain in the red vote in the $i^{th}$ precinct in the ordering is then $redSLOPE*(b_i+r_i – 800)*(b_i+r_i)$, and the total gain in red votes (in terms of the linear regression) in all blue win precincts $\geq 800$ is

$$redLPB = \sum_{i=N^*}^{N} redSLOPE * (b_i + r_i - 800)(b_i + r_i).$$

*redLPB* is an estimate of the total red vote gain (blue vote gain) when *redSLOPE* > 0 (< 0) due to LPB in blue win precincts with vote total $\geq 800$. If *redSLOPE* < 0, then the gain of blue votes in the $i^{th}$ precinct in the ordering is $blueSLOPE*(b_i+r_i – 800)*(b_i+r_i)$, where *blueSLOPE* = – *redSLOPE*, and the total blue vote gain is

$$blueLPB = \sum_{i=N*}^{N} blueSLOPE * (b_i + r_i - 800)(b_i + r_i).$$

A similar framework is carried out for the red win precincts.

## 4. The Example of Pennsylvania

An example of LPB is given for Pennsylvania in 2008. There were 5,903,655 votes in 9,241 precincts, 3,260,761 blue votes (55.2%) and 2,642,894 red votes (44.8%), as reported in [Harvard Election Data Archive] (other sources report slightly different totals). The blue win precincts totaled 5,482 and the red win precincts 3,736 (21 precincts had exactly the same total votes and are not counted). Consider first the blue win precincts. The 5,482 blue win precincts had a total of 3,228,368 votes, 2,178,735 blue votes, and 1,049,633 red votes. To motivate the analysis divide the blue win precincts into 17 precinct size brackets, in increasing intervals of 200: bracket 1 (1-199), bracket 2 (200-399), bracket 3 (400-599), *etc*. Figure 1 shows the number of precincts and votes in the 17 brackets, and Figure 2 shows the means of the blue and red vote fractions in each of the seventeen brackets.

     A pattern of increase in red fractions and decrease in blue fractions (mirrored about 0.5) is evident in Figure 2, particularly in brackets > 4 (precincts with total votes $\geq$ 800). Further, the increase appears approximately linear, which indicates a correlative dependence on precinct size. This linear correlation supports the validity of a least squares linear regression analysis for the set of blue win precincts with size $\geq$ 800, as in the framework (1) – (3) above. In Figure 3 a scatter plot is given for the blue and red fractions of all 5,482 blue win precincts in increasing size order. The vertical lines represent the bracket boundaries as above, with the green vertical line separating precincts with size less than and greater than 800. The blue and red lines represent the linear regression curves for the blue and red vote fractions in blue win precincts with size $\geq$ 800, and clearly show a significant increase in the red fractions and complementary decrease in blue fractions.

     A similar analysis is given for the red win precincts for Pennsylvania in 2008. There were 2,662,205 votes in 3,736 red win precincts, with 1,586,720 red votes and 1,075,485 blue votes. As with the blue win precincts, the red win precincts are divided into 18 size brackets in increasing intervals of 200. Figure 4 shows the number of precincts and votes in the 18 brackets, and Figure 5 shows the means of the blue and red vote fractions in the 18 brackets. In Figure 6 a scatter plot is given for the blue and red fractions of all 3,736 red win precincts in order of increasing size. As with the blue win precincts, a least squares linear regression analysis for the set of red win precincts with size $\geq$ 800 is carried out. The blue and red lines represent the linear regression curves for the blue and red vote fractions in red win precincts with size $\geq$ 800, showing no significant change in the red and blue fractions. In Figure 7 scatter plots are given for a typical random ordering of the 5,482 blue win precincts and a typical random ordering of the 3,736 red win precincts in Pennsylvania in 2008. No pattern is evident in either, in contrast to the significant LPB for red votes in blue win precincts $\geq$ 800 in Figure 3.

     The analysis of LPB in precincts $\geq$ 800 in Pennsylvania in 2008 reveals the presence of significant red LPB in blue win precincts $\geq$ 800 and the absence of significant blue or red LPB in red win precincts $\geq$ 800. If only the votes in all precincts $\geq$ 800 were counted, the Republican candidate would have won the state (1,473,868 red votes - 50.5%, 1,447,055

blue votes - 49.5%). At issue are the contributions of voter heterogeneity and voter inconvenience to LPB in precincts $\geq$ 800. In red win precincts $\geq$ 800, the absence of blue LPB suggests that voter heterogeneity is either absent or offset by blue voter inconvenience. Blue win and red win precincts differ demographically, but if voter heterogeneity is less significant in red win precincts $\geq$ 800, it is arguable that it is less significant in blue win precincts $\geq$ 800. Thus, the striking red LPB in blue win precincts, with outcome opposite to the statewide outcome, is due largely to voter inconvenience of blue voters.

**5. Analysis of LPB in 20 States**

As in the example of Pennsylvania 2008, the framework (1) – (3) is applied to 19 additional states (Figures 8,9,10, Table 1, and Supporting Information available on request). Precincts are ordered according to increasing total vote size and least squares linear regression curves (blue for Democrat fractions and red for Republican fractions) are fitted to precincts with size $\geq$ 800. As in the example of Pennsylvania, if the red fractions are increasing, then the slope of the linear regression curve is *redSLOPE* > 0, and the red vote gain = *redLPB* votes = the blue vote loss. For each state the ratio of *redLPB* or *blueLPB* votes to total votes cast in all precincts in the state is given as a percent. A summary is given in Table 1.

    Table 1 reveals patterns in LPB for the 20 states analyzed. In 11 states LPB is relatively small (< 0.5%) in both blue win and red win precincts (Arizona, Colorado, Idaho, Indiana, Kansas, Louisiana, Mississippi, Nevada, Ohio, Tennessee, Washington). In 9 states LPB is relatively large ($\geq$ 0.5%) in blue win precincts (Alabama, Delaware, Florida, Michigan, Missouri, New Mexico, North Carolina, Pennsylvania, Wisconsin), and in all of these states the LPB in blue win precincts is red. In these 9 states the corresponding LPB in red win precincts is relatively small (< 0.5%) in all but Michigan, New Mexico, and North Carolina. In red win precincts $\geq$ 800 in Michigan 2012 (Figure 8) LPB is red and *redLPB* = 0.63% of state votes cast, which indicates that voter inconvenience also penalizes blue voters in red win precincts $\geq$ 800. In New Mexico (Supporting Information Figure S13) and North Carolina (Figure 9) the LPB in red win precincts $\geq$ 800 is blue, which indicates that voter heterogeneity in large precincts is significant in red win precincts $\geq$ 800. If New Mexico and North Carolina are viewed as exceptional cases, then Alabama, Delaware, Florida, Michigan, Missouri, Pennsylvania, and Wisconsin have significant asymmetry in LPB, namely, a significant presence of red LPB in blue win precincts and a significant absence of blue LPB in red win precincts. This asymmetry is indicative of significant blue voter inconvenience in large precincts in these states

**6. Voter Heterogeneity and Voter Inconvenience in LPB**

In general, voter heterogeneity and voter inconvenience in US presidential election outcomes involve multiple factors, which are difficult to separate. Two examples illustrate the extremes: North Carolina 2008 and Ohio 2008.

In North Carolina 2008 the precincts are very large, and red LPB is very significant in blue win precincts $\geq$ 800, as is blue LPB in red win precincts $\geq$ 800 (Figure 9). Further, there is a balance of the two. If an analysis is carried out for combined blue win an red win precincts in North Carolina, and linear regression is fitted to the blue and red fractions in all precincts $\geq$ 800, both blue win and red win, neither red LPB nor blue LPB appear. In North Carolina 2008 it is not possible to evaluate the role of voter inconvenience as a function of precinct size, although it may be present and disguised by the very large size of almost all precincts in the state.

In Ohio 2008 the precincts are very small, and there is neither red LPB nor blue LPB in precincts $\geq$ 800 (Figure 10). If the analysis is carried out for precincts $\geq$ 600 instead of precincts $\geq$ 800, *redLPB* in blue win precincts = 0.19% of total state votes cast and *blueLPB* in red win precincts = 0.09% of total state votes cast, both still insignificant. In Ohio 2008 the analysis does not reveal LPB dependent on voter inconvenience. It is relevant, however, that Ohio has nearly twice as many precincts as polling places, which means that polling places could be biased by voter inconvenience as polling place size increases.

To identify the role of voter inconvenience in LPB it is useful to consider vote outcomes in counties as well as states, particularly large urban counties with many precincts and strong Democratic margins in votes cast. In such counties demographic diversity in precincts is reduced in comparison to demographic diversity statewide. It is also useful to consider a lower value than 800 for the designation of a large precinct. Two counties are illustrated for the presidential election in 2012: Figure 11 - Dade County (Miami) in Florida and Figure 12 - Wayne County (Detroit) in Michigan. The LPB percentage of total county votes is computed both for precincts $\geq$ 800 and precincts $\geq$ 600. Both show significant *redLPB* percentage in blue win precincts $\geq$ 800 and even greater *redLPB* in blue win precincts $\geq$ 600. Both also show no significant presence of *blueLPB* in red win precincts $\geq$ 600 or $\geq$ 800, and in the case of Wayne county positive *redLPB* in red win precincts. The analysis quantifies a significant blue vote loss to the Democratic presidential candidate in these two counties due to voter inconvenience. Both counties experienced long lines at some polling places (http://www.miamiherald.com/2012/11/06/3085322/across-south-florida-long-lines.html, and http://detroit.cbslocal.com/2012/11/06/heavy-election-turnout-brings-lines-ballot-shortage/).

## 7. Discussion

In US history voter suppression has taken many forms, which over time have been recognized and resolved. Currently, there is a less recognized, but significant form of voter suppression throughout many parts of the US, which is identified here as voter inconvenience. Voter inconvenience may disenfranchise any voter subjected to it, but takes a heavier toll on Democratic candidates in US presidential elections. The resolution of voter suppression due to voter inconvenience requires multiple efforts of government and election administrations. The US Election Assistance Commission and the Brennan

Center for Justice, New York University School of Law have provided extensive recommendations to improve efficiency and fairness in US elections, and many of these relate to voter inconvenience.

Important issues are the distribution of voting machines to precincts (1), voting queues (2), (3), waiting times (4), adequate early voting periods (5), accessible polling facilities (6), (7), and simplified ballots (Center for Civic Design, http://centerforcivicdesign.org). All impact voter inconvenience, and all require recommendations for improvement. The analysis here supports another recommendation to resolve voter suppression due to voter inconvenience. Every state and every county in the US should be statutorily mandated to provide complete precinct level reports for elections. These precinct reports should include early, absentee, election day, provisional, and rejected ballot counts, number of polling places, number of voting machines, number of poll workers, voter arrival rates, extended voting hours, as well as information about excessively long queues, excessively long voting times, and voting machine dysfunction. The transparency and validity of the democratic process in US elections would be greatly strengthened if every precinct in the US were held accountable for its fundamental electoral function.

**Figure Legends**

Figure 1. Precinct numbers and total votes in the 17 brackets of increasing intervals of 200 for the 5,482 blue win precincts in Pennsylvania in 2008.

Figure 2. Means for the blue and red vote fractions for the 17 brackets of blue win precincts ordered by total vote size in increasing intervals of 200 for Pennsylvania in 2008. The bar width of each bracket is proportional to the ratio of total blue win votes in that bracket to the total blue win votes in all blue win precincts.

Figure 3. Blue win precincts scatter plot and linear regression curves for precincts with size $\geq$ 800 for Pennsylvania in 2008. The slope of the red linear regression is *redSLOPE* = 0.00005 = – the slope of the blue linear regression. The total red vote gain = *redLPB* = 33,466 = 0.57% of the total votes cast in the state in all precincts.

Figure 4. Precinct numbers and total votes in 18 brackets of increasing intervals of 200 for the 3,736 red win precincts in Pennsylvania in 2008.

Figure 5. Means for the blue and red vote fractions for the 18 brackets of red win precincts ordered by total vote size in increasing intervals of 200 for Pennsylvania in 2008. The bar width of each bracket is proportional to the ratio of total red win votes in that bracket to the total red win votes in all red win precincts.

Figure 6. Red win precincts scatter plot and linear regression curves for precincts with size $\geq$ 800 for Pennsylvania in 2008. The slope of the blue linear regression is *blueSLOPE* = 0.00000054 = – the slope of the red linear regression. The total blue votes gained = *blueLPB* = 508 = 0.0086% of the total votes cast in the state in all precincts.

Figure 7. Left panel: Blue win precincts scatter plot and linear regression curves for the 5,472 blue win precincts randomly ordered for Pennsylvania in 2008. Right panel: Red win precincts scatter plot and linear regression curves for the 3,736 red win precincts randomly ordered for Pennsylvania in 2008. The linear regression slopes are all approximately 0.

Figure 8. Michigan 2012. Left panel: 2975 blue win precincts, blue votes = 1,642,558, red votes = 754,550, *redSLOPE* = .000079, *redLPB* = 67,008 red vote gain = 1.4% of state votes. Right panel: 2,510 red win precincts, blue votes = 1,020,426, red votes = 2,488,167, *redSLOPE* = .000028, *redLPB* 30,995 red vote gain = 0.63% of state votes.

Figure 9. North Carolina 2008. Left panel: 1105 blue win precincts, blue votes = 1,227,437, red votes = 565,982, *redSLOPE* = .000015, *redLPB* = 40,032 red vote gain = 0.94% of state votes. Right panel: 1,585 red win precincts, blue votes = 905,367, red votes = 1,557,163, *blueSLOPE* = .000015, *blueLPB* = 50,105 blue vote gain = 1.2% of state votes.

Figure 10. Ohio 2012. Left panel: 5,617 blue win precincts, blue votes = 1,794061, red votes = 862,780, *redSLOPE* = .0000057, *redLPB* = 174 red vote gain = 0.0031% of state votes. Right panel: 5,472 red win precincts, blue votes = 1,141,941, red votes = 1,811,214, *blueSLOPE* = .000015, *blueLPB* = 912 blue vote gain = 0.016% of state votes.

Figure 11. Dade County 2012. Left panel: 569 blue win precincts, blue votes = 206,125, red votes = 87,266, *redLPB* = 1.1% of county votes for precincts $\geq$ 800 and 3.0% for precincts $\geq$ 600. Right panel: 214 red win precincts, blue votes = 44,304, red votes = 61,525, *redLPB* = 0.000090% of county votes for precincts $\geq$ 800 and 0.012% for precincts $\geq$ 600.

Figure 12. Wayne County 2012. Left panel: 1,109 blue win precincts, blue votes = 533,660, red votes 134,771, *redLPB* = 1.4% of county votes for precincts $\geq$ 800 and 3.2% for precincts $\geq$ 600. Right panel: 113 red win precincts, blue votes = 56,487, red votes = 74,426, *redLPB* = 0.27% of county votes for precincts $\geq$ 800 and 0.14% for precincts $\geq$ 600.

**Figure 1**

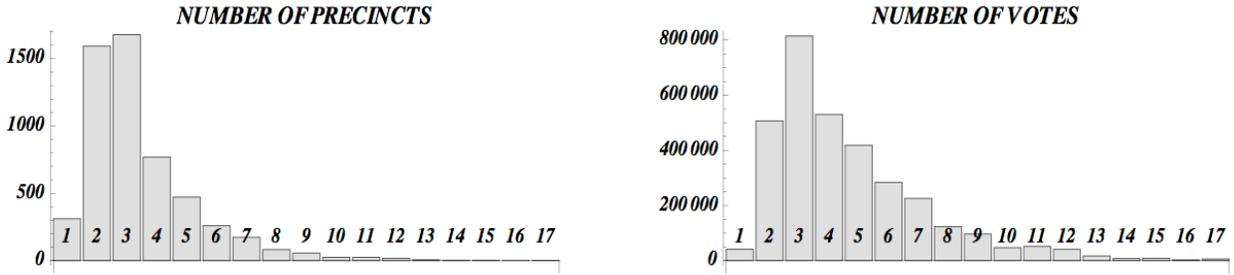

**Figure 2**

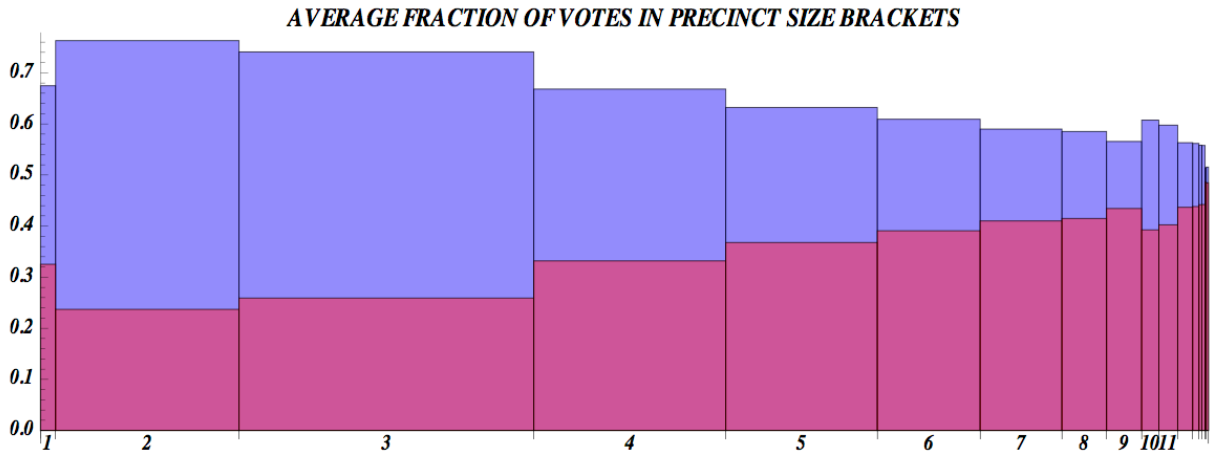

**Figure 3**

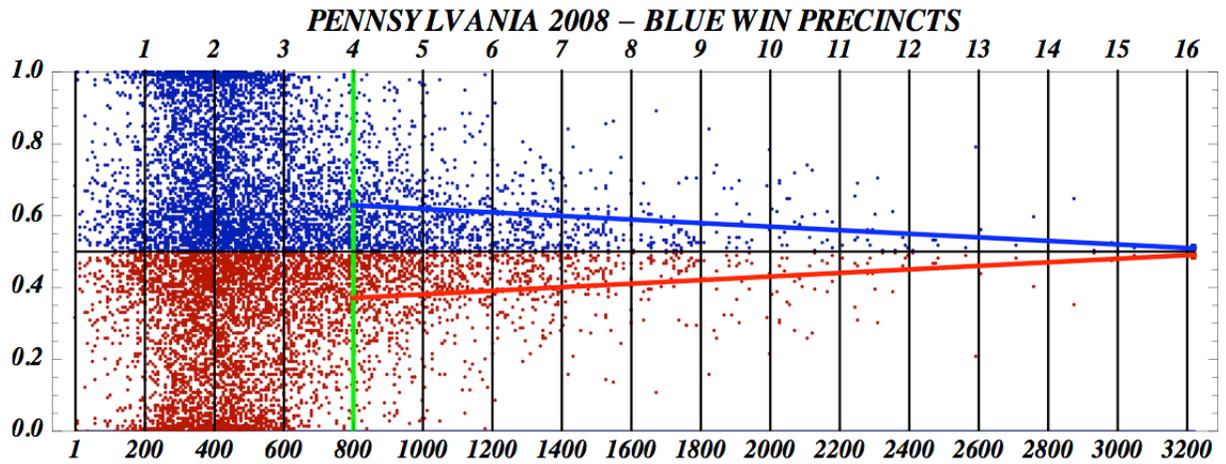

**Figure 4**

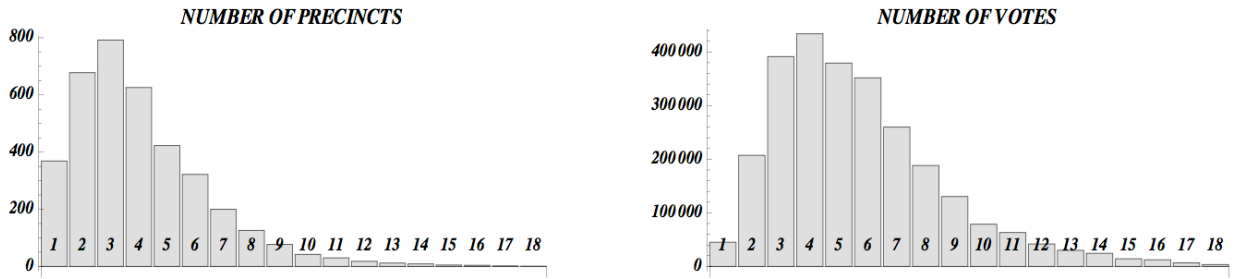

**Figure 5**

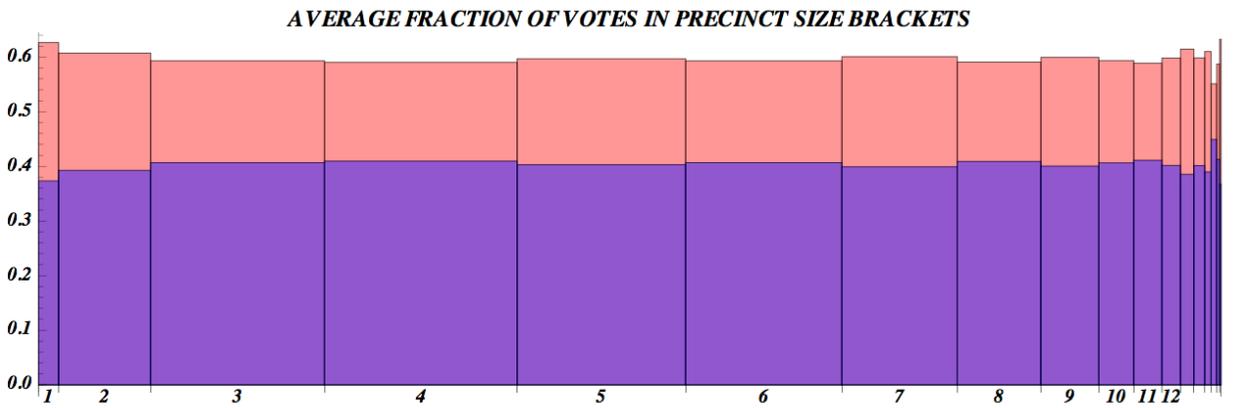

**Figure 6**

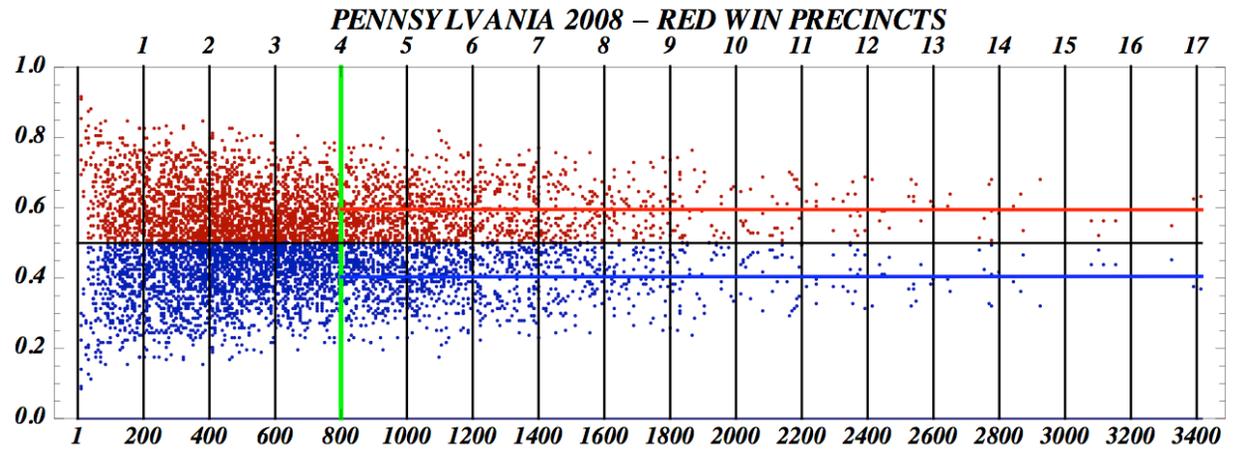

**Figure 7**

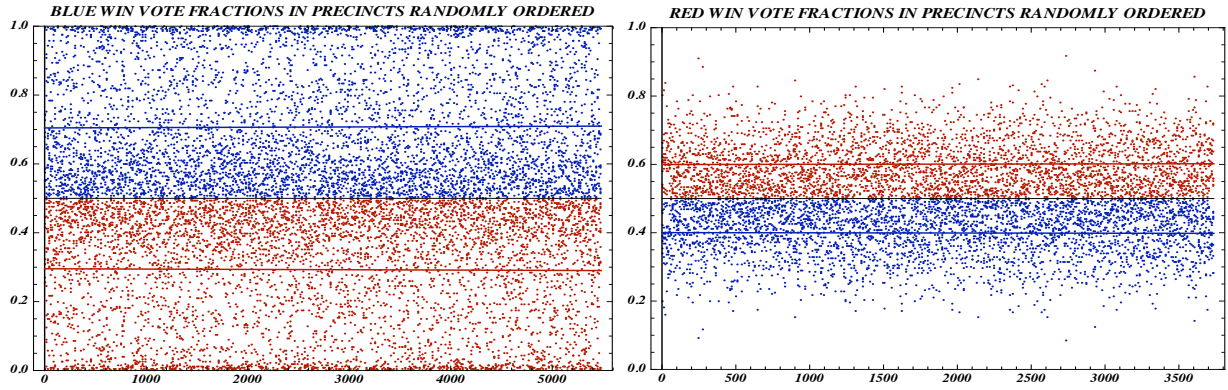

**Figure 8**

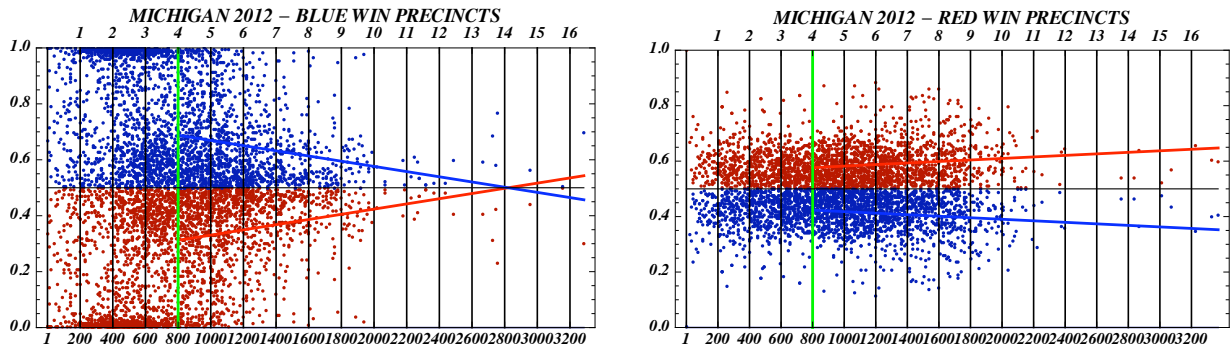

**Figure 9**

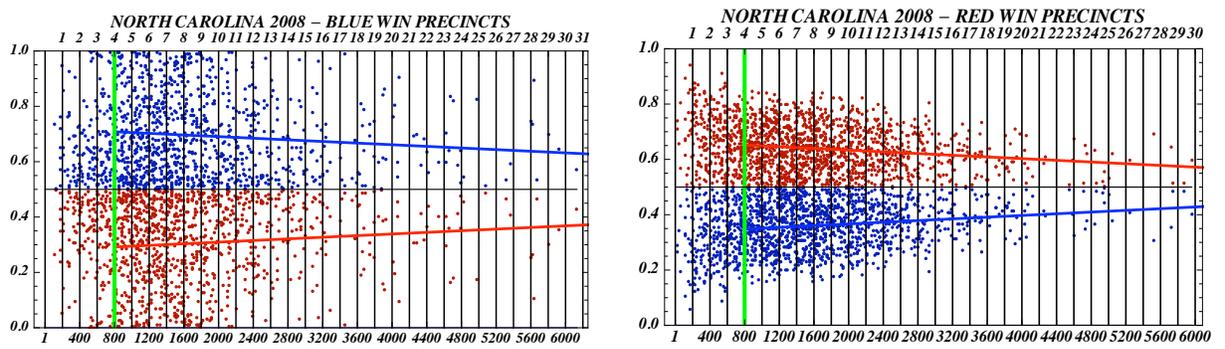

**Figure 10**

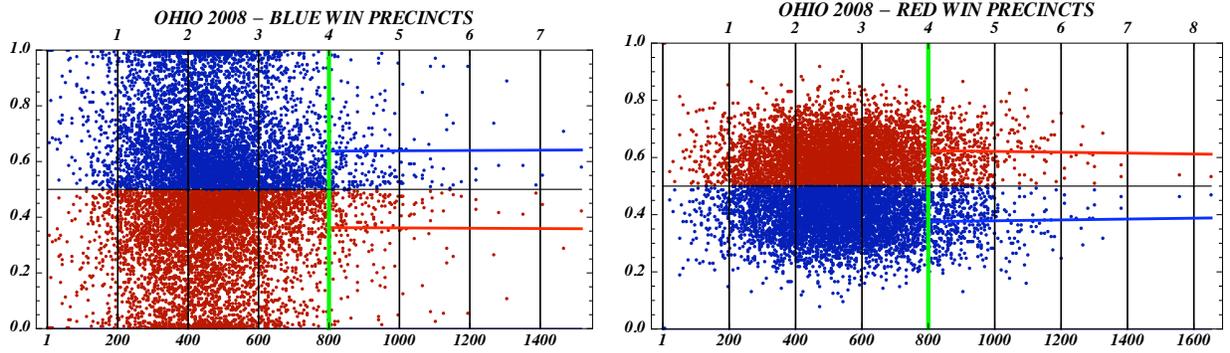

**Figure 11**

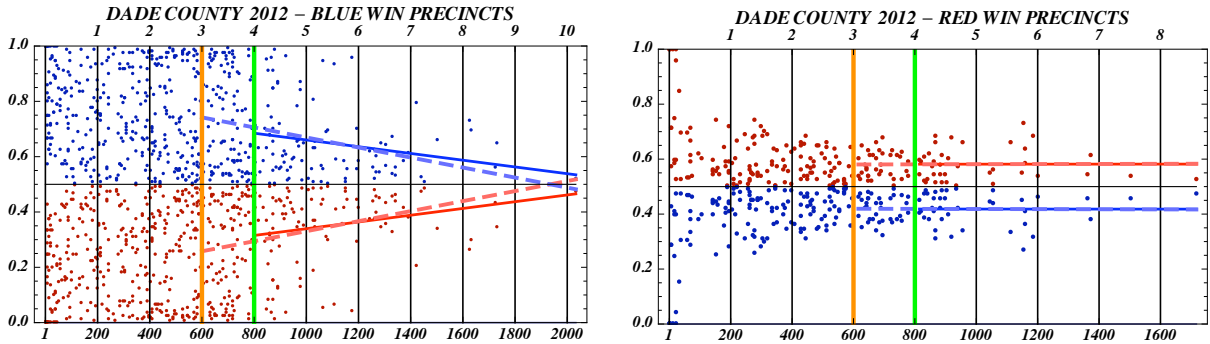

**Figure 12**

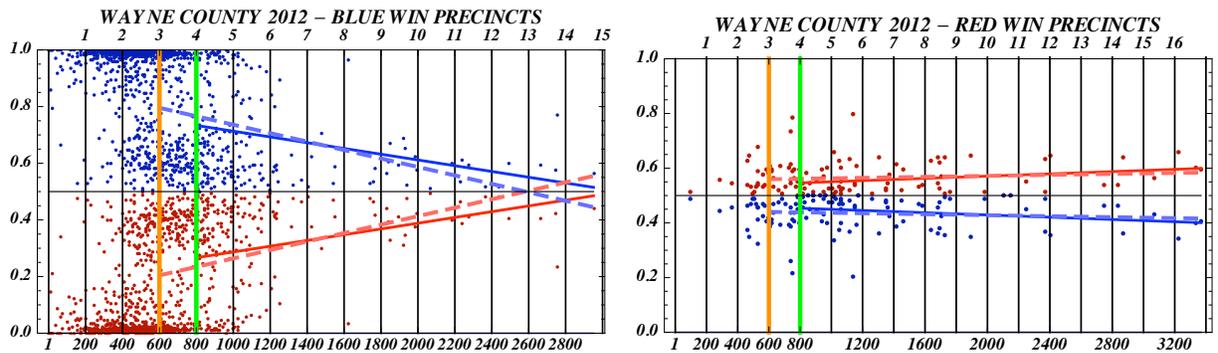

**Table 1**

**Summary of the large precinct bias for states in precincts with total vote ≥ 800**

| State - year | Blue win precincts | | Red win precincts | |
|---|---|---|---|---|
| | Mean precinct size | blue or red LPB % of total state votes | Mean precinct size | blue or red LPB % of total state votes |
| Alabama 2008 | 734 | 0.53% | 793 | 0.075% |
| Arizona 2008 | 797 | 0.28% | 1,157 | 0.38% |
| Colorado 2008 | 716 | 0.42% | 748 | 0.28% |
| Delaware 2008 | 900 | 3.1% | 1142 | 0.20% |
| Florida 2008 | 1,106 | 1.1% | 1,310 | 0.34% |
| Idaho 2008 | 780 | 0.065% | 576 | 0.29% |
| Indiana 2008 | 485 | 0.030% | 322 | 0.063% |
| Kansas 2008 | 630 | 0.22% | 322 | 0.24% |
| Louisiana 2008 | 389 | 0.17% | 471 | 0.025% |
| Michigan 2012 | 806 | 1.4% | 991 | 0.63% |
| Mississippi 2008 | 606 | 0.18% | 675 | 0.30% |
| Missouri 2008 | 530 | 0.75% | 616 | 0.079% |
| Nevada 2008 | 631 | 0.28% | 448 | 0.14% |
| New Mexico 2008 | 573 | 0.50% | 568 | 0.63% |
| North Carolina 2008 | 1,623 | 0.94% | 1,554 | 1.2% |
| Ohio 2012 | 575 | 0.0013% | 610 | 0.016% |
| Pennsylvania 2008 | 589 | 0.57% | 713 | 0.0086% |
| Tennessee 2008 | 1,201 | 0.24% | 1,093 | 0.055% |
| Washington 2008 | 452 | 0.26% | 430 | 0.35% |
| Wisconsin 2008 | 824 | 1.2% | 918 | 0.12% |